# 'Cored Apple' Bipolarity : A Global Instability to Convection in Radial Accretion?

R.N. Henriksen[1] and D. Valls–Gabaud[1,2]★

[1] *Astronomy Group, Stirling Hall, Queen's University, Kingston, Ontario K7L 3N6, Canada*
[2] *Institut d'Astrophysique de Paris, 98 bis, Bld. Arago, 75014 Paris, France*

**ABSTRACT**
We propose that the prevalence of bipolarity in Young Stellar Objects ( YSO's) is due to the fine tuning that is required for spherical accretion of an ambient medium onto a central node. It is shown that there are two steady modes that are more likely than radial accretion, each of which is associated with a hyperbolic central point in the meridional stream lines, and consequently with either an equatorial inflow and an axial ejection or *vice versa*. In each case the stream lines pass through a thick accretion torus, which is better thought of as a standing pressure wave rather than as a relatively inert Keplerian structure. We base our arguments on a simple analytic example, which is topologically generic, wherein each bipolar mode is created by the 'rebound' of accreting matter under the action of the thermal, magnetic, turbulent and centrifugal pressures created in the flow. In both bipolar modes the presence of non-zero angular momentum implies axial regions wherein the pressure is first reduced below the value at infinity and then becomes negative, where the solution fails because rotating material can not enter this region without 'suction'. The models thus have empty 'stems' where the activity of the central source must dominate. So the basic engine of the bipolar flow discussed here is simply the rebound of freely falling material from a thick pressure disc into an axial low pressure region. The low mass, high velocity outflow must be produced in this region by an additional mechanism. This is reminiscent of the 'cored apple' structure observed recently in the very young bipolar source VLA 1623.

## 1 INTRODUCTION

Bipolar outflows are widespread among YSOs and active galactic nuclei (AGN), and it is dangerous to presume that one model can apply to all objects or even to all evolutionary stages of any given object. However it might be possible to discover a certain universality in a set of minimum requirements for bipolarity, which would be relatively independent of details of the central source such as luminosity or spectral appearance. One could expect such minimum conditions to appear in the limit where well developed large-scale bipolarity is already present around very cold, low luminosity central sources. Thus it is to the latter class of objects that we direct the model of this paper.

Although such objects may well exist among galactic-scale objects, and indeed the non-jet form of the solution below represents an interesting example of accretion onto a disc and/or active galactic nuclei, we are inspired especially by the youngest protostellar outflow sources such as the recently discovered object VLA 1623 in $\rho$ Ophiuchi A (André et al. 1990). The source, which is only detected at centimetric and (sub ) millimetric radio wavelengths, drives a well collimated CO bipolar outflow . The mechanical luminosity of the outflow is $\approx$ 0.5- 2 $L_\odot$, about the same as that of the prototypical source L1551 (ibid).

Sub-millimetric observations (André, Ward–Thompson and Barsony, 1993) have established somewhat unexpectedly that VLA 1623 is a low-luminosity ($L_{bol} \leq L_\odot$) YSO surrounded by a cold ($T \leq 20$ K), compact ($R \approx 1000 AU$), and relatively massive ($M_c \approx 0.6 M_\odot$) circumstellar structure *in a nearly spherically symmetric distribution*, with a very weak density gradient ($\rho \propto r^{-1/2}$). The authors recognize various difficulties in this discovery for the large-disc based models of bipolar outflows (e.g. Pelletier and Pudritz 1992; Königl 1989; Blandford and Payne 1982), and for those models which identify such sources with continuing accretion onto a central source (Choe and Henriksen 1986; Shu, Adams and Lizano 1987) since the object is so sub-luminous. The authors also suggest a graphically succinct description of such objects as 'cored apples', where the bipolar flow presumably coincides with, or emerges from, the removed core.

In this paper we suggest a basis for understanding such bipolar sources based on the exact, global self-similar solutions found by Wang (1983) and elaborated by Henriksen (1987 $\equiv$ H87). The solution is also known independently in an experimental context (Goldshtik 1990, Squire 1952). Briefly, we will describe them as rising axial jets in an axially-symmetric, dipolar convection pattern forming a protostellar extended 'atmosphere'. The circulation is driven by dynamically maintained overpressures which arise in a choked stationary accretion flow.

These models are very much in the spirit of the prescient paper by Lynden–Bell (1978) where it was shown that 'double whirlpools' could exist in a rotating gaseous 'atmosphere' around a gravitating point mass. The present models allow for this and also for meridional circulation, and they are pressure-gradient driven. In this latter respect they are similar to the recent models of Lovelace, Berk and Contopoulos (1991), but they differ in that they are global solutions which automatically contain 'discs' and regions of axial collimation (see also Igumenshchev et al. 1993, for another interesting case). Only the physics of a moderate heating in the jet region near the core needs to be added in the present models. A subsequent paper will present a study of the more realistic cases that we have studied numerically. The analytical models presented here should *not* be considered as the best that this type of model can achieve. Nevertheless they do demonstrate that bipolar outflow can exist in the absence of a large scale thin disc structure.

## 2   THE ANALYTIC MODEL

The solutions are the two-sided generalizations of the Landau-Squires submerged jet solution. The model chosen has a uniform density, a constant Alfvén number, a constant *specific* angular momentum, and a constant viscosity which may be zero. The existence of either of these latter two constants constrains the velocity to decrease as the reciprocal radius. The magnetic field is everywhere parallel or anti-parallel to the stream lines which, in a true bipolar outflow solution (but see also the axial accretion case below), have meridional projections that approach parallel to the equatorial plane near this plane and then deflect to become parallel to the axis of rotation near this axis. They also rotate about this axis of course. This latter arrangement is a simple way to ensure stationary hydromagnetic flow, but it overlooks a thin equatorial region of dissipation and/or ambipolar diffusion that one may expect to be present in general. It also excludes models in which significant energy is transported by the Poynting flux. Such a region is important for the bipolar flow models in which an equatorial disc drives the outflow by twisting the field (see e.g. Königl 1989; Wang, Lovelace and Sulkanen 1990; Pelletier and Pudritz 1992) but here, since we are primarily interested in the large-scale circulation pattern established by pressure gradients, it is of less importance. In fact the active discs often discussed in these contexts are likely to be on a much smaller spatial scale, being in effect a subscale boundary layer between the environment and the star.

There is moreover no net accretion onto the central object if its finite radius is neglected, which is how these models avoid the excess luminosity problem and remain strictly stationary. In fact our entire flow region is an extended boundary region between the inner core-subscale disc and the outer cloud material and may be considered as a protostellar 'atmosphere'. The image of an atmospheric circulation is apt for this analytic case since the velocity decreases more rapidly than does the free fall speed at infinity.

The relevant solution is of a self-similar form which is expressible as (H87):

$$
\begin{aligned}
\mathbf{v} &= (\nu/a)\, r^{-1}\, \mathbf{U}(\theta)\,, \\
\mathbf{B} &= \sqrt{4\pi\rho}\, (\nu/a)\, r^{-1}\, \mathbf{U}(\theta)/M_A\,, \\
p &= p_\infty + \frac{\rho\gamma}{r} + \left(\frac{\rho\nu^2}{ar^2}\right)\left(P(\theta) - \frac{|\mathbf{U}|^2}{2(M_A^2-1)}\right)\,.
\end{aligned} \quad (1)
$$

Here $\nu$ is the constant viscosity, $M_A$ is the Alfvénic number, $a \equiv 1 - M_A^{-2}$, $\gamma \equiv GM_*$ for some central mass $M_*$ and Newton's constant $G$, $p_\infty$ is any non-zero pressure at spatial infinity, and other symbols have their usual meanings as in H87. The second term in the pressure is a spherically symmetric 'virial' term, while the third term is the dynamic pressure.

The analytic solution takes the explicit form for super-Alfvénic flow ($a > 0$);

$$U_\theta = f(x)\sqrt{1-x^2}\,, \quad (2)$$

$$U_r = -2x f(x) - \frac{(1-x^2)}{2} f^2(x) - \frac{\sigma_+}{2} \cdot \frac{1-x}{1+x} \quad (3)$$

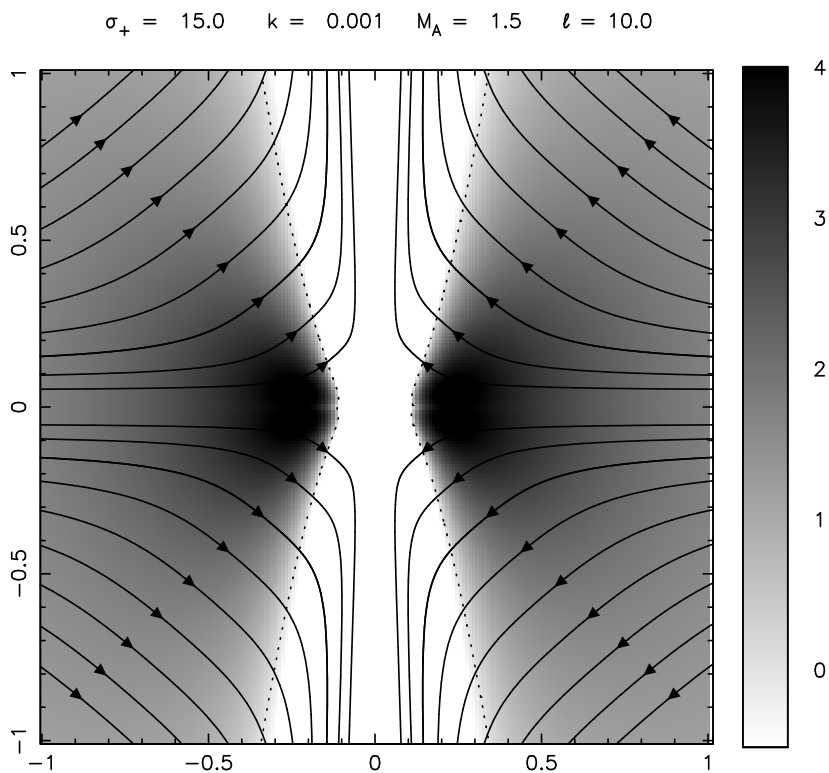

**Figure 1.** Bipolar *outflow* mode in an extremely collimated limit, where $\sigma_+ = 15.0$, $\ell = 10.0$, $M_A = 1.5$, and $k = 0.001$. The meridional stream-field lines are shown superimposed on the grey-coded contours of the scaled pressure difference $\Delta$ (see Eq. 15). The dotted lines give the meridional section of the surface $\Delta = 0$.

The axis of symmetry of the figure is the axis of rotation.

$$U_\phi = \ell/\sqrt{1-x^2}, \qquad (4)$$

$$P(x) = \frac{\sigma_+}{2} - \frac{U_\theta^2 + U_\phi^2}{2} + [U_r]. \qquad (5)$$

In these expressions, $x \equiv \cos\theta$, and so is zero on the equator, $f(x)$ an angular function described below, $\ell$ is the constant specific angular momentum, and $\sigma_+$ is a constant such that the sum of the transverse stresses $T_{\theta\theta} + T_{\phi\phi}$ is given by

$$T_{\theta\theta} + T_{\phi\phi} = (\sigma_+)(\rho v^2/a)r^{-2} + 2p_\infty. \qquad (6)$$

The expressions for the remaining stress components are available in H87, and the solution has been written in the form appropriate to the domain $0 \leq \theta \leq \pi/2$. The solution is extended continuously into $\pi/2 \leq \theta \leq \pi$, by a mirror reflection in the equatorial plane, and an axial rotation extends the solution to the whole space. This symmetry ensures that $U_\theta$ and consequently $B_\theta$ vanish at the equator. The tangential velocity is parallel on both sides of the equator, but it is more in accordance with possible magnetic fields accreted from infinity in super-Alfvénic flow if the field (i.e. $M_A$) changes sign across the equator.

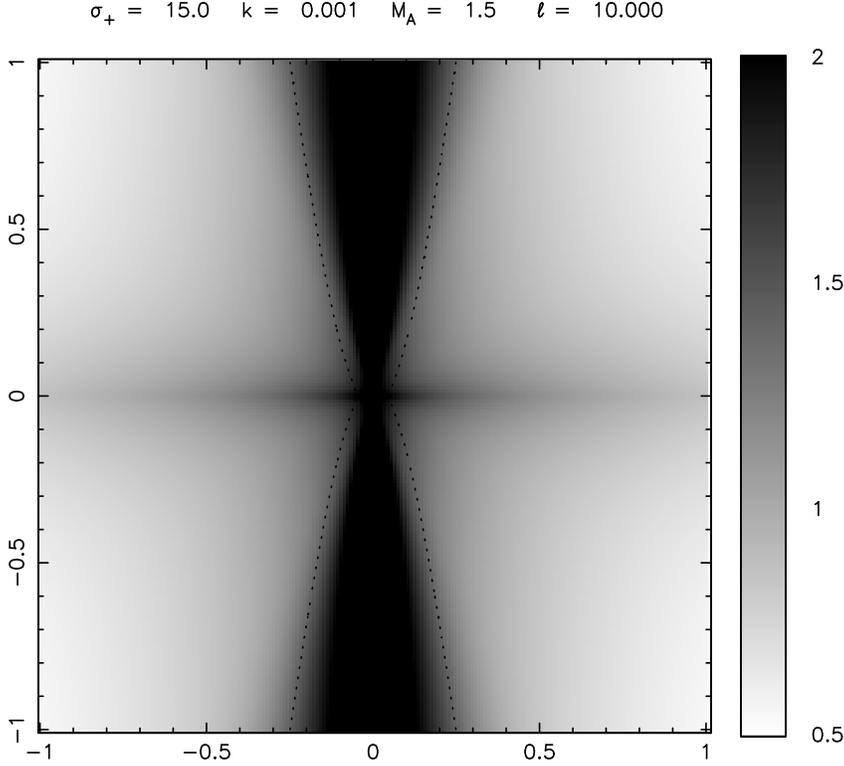

**Figure 2.** Grey-coded contours of $\log_{10}\left(v_r^2 + v_\theta^2\right)^{1/2}$ for the outflow mode of Fig. 1.

The non-singular meridional projections of the stream-magnetic field lines are given by $rU_\theta \sin\theta = constant$, or in explicit form

$$f(x)(1 - x^2) = -s/r ,  \qquad (7)$$

where $s$ is a stream line constant. The flow field is now evidently self-similar since as $s$ is varied only the scale of $r(x)$ changes, and not the functional form.† In addition to this family of stream lines there are singular envelope curves $U_\theta = 0$, $d\theta = 0$, which lie along $\theta = 0$ (polar axis) and $\theta = \pi/2$ in the meridional plane.

When $\ell$ is not zero, another boundary region arises along the polar axis of the flow. For at some polar angle sufficiently small but dependent on radius in general, the pressure will become negative after first becoming less than the value at infinity (see Figures 1–3). Material ejected from the central object has thus a preferred route along the 'stem' of the 'apple'.

We must now specify the angular function $f(x)$. The mirror reflection symmetry referred to above [i.e. if $f(x, \sigma_+)$ is a solution then so is $-f(-x, -\sigma_+)$] permits us to concentrate on the form in $[0, \pi/2]$ with the appropriate boundary condition at the equator. Requiring $U_\theta(x = 0) = 0$, we have for $\sigma_+ > 1$ and $\chi = \ln(1+x)$

---

† The solution remains unchanged in the absence of viscosity, except for the last term (placed in square brackets) of equation (5) which must be suppressed in that case.

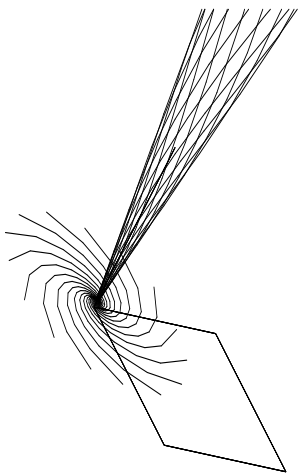

**Figure 3.** Three dimensional representations of the stream lines for the case described in Figs. 1 and 2.

$$f(x) = -\sigma_+ e^{-\chi} \left( \frac{\tan\left[\frac{\sqrt{\sigma_+ - 1}\,\chi}{2}\right]}{\sqrt{\sigma_+ - 1} - \tan\left[\frac{\sqrt{\sigma_+ - 1}\,\chi}{2}\right]} \right) . \tag{8}$$

We remark immediately that the nature of the solution is rather sensitive to the value of $\sigma_+$. Sufficiently large values will give unphysical singular velocities at some $x < 1$, and the inf of this singular domain of $\sigma_+$ is given by the first value that creates a physical singularity at $x = 1$, that is $\sigma_+ \approx 15.3455$. The flow has a central unstable fixed point called a hyperbolic or saddle point for $\sigma_+ \in [0, 15.35[$, and an example is shown in Figures 1 to 3. This topology is continuous across $\sigma_+ = 1$, since at precisely this value we have (H87)

$$f(x) = -(1+x)^{-1} \frac{\ln(1+x)}{2 - \ln(1+x)} , \tag{9}$$

which defines a flow field having a hyperbolic fixed central point.

Moreover for $\sigma_+ < 1$, although the form of the solution abruptly changes to

$$f(x) = 2p_1(1+x)^{-1} \left( \frac{1 - (1+x)^{(p_2 - p_1)}}{1 - \frac{p_1}{p_2}(1+x)^{(p_2 - p_1)}} \right) , \tag{10}$$

where

$$p_{1,2} = \frac{1 \pm \sqrt{1 - \sigma_+}}{2} , \tag{11}$$

we can readily establish that for $p_2 > 0$, which holds when $\sigma_+ > 0$, the fixed point at the centre remains hyperbolic. At a given radius, the ratio of the radial velocity near the axis of rotation to that at in the equatorial plane decreases with $\sigma_+$ to a value near unity when $\sigma_+ = 1$, and remains near this value thereafter.

The interesting development occurs at $\sigma_+ = 0$ however. In this case there is no solution that satisfies the reflection symmetry condition ($U_\theta(\pi/2) = 0$) unless $U_\theta$ vanishes everywhere. But in such a case there is no way to satisfy continuity within the restriction of our assumed self-similar form. In fact this condition of zero transverse stress (relative to the value at infinity) corresponds to the Bondi (1952) radial compressible steady accretion, or to the artificially stress-balanced flow with non-zero specific angular momentum (Henriksen and Heaton, 1975). We are thus able to regard these cases as rather special members of a much more general ensemble.

As $\sigma_+$ becomes negative so that $p_2$ is also negative, we see from equation (10) that $f$ is always positive for $\theta < \pi/2$ and from equations (2) and (3) that *there has been a remarkable reversal in the sense of the hyperbolic flow*. The accretion is now bipolar *along* the rotation axis and the ejection is now an axially symmetric rotating outflow that is symmetric about the equatorial plane. The 'cored apple' is now the pressure toroid (see Figures 4 to 6) that is established by the convection flow as it descends parallel to the stem and emerges parallel to the equatorial plane. This flow is driven by the pressure difference between infinity and a rarefied central 'cavity'. In fact for $\sigma_+$ less than about $-10$ we find that the pressure in this mode must be everywhere less than that at infinity. Thus it would only arise around an object embedded in sufficiently high pressure environments.

Interestingly enough this latter mode reproduces in broad outline the type of accretion-driven dynamical thick disc that was found to exist (behind an outward propagating wave front) in numerical simulations carried out by Clarke, Karpik, and Henriksen (1985), and also by Loken (1986) in an extensive series of pertinent calculations. An example of the properties of this mode in the present analytic context is shown in Figures 4 to 6.

The dimensionless transverse stress is a physical quantity that is in general comprised of viscous stress, as well as magnetic, thermal, dynamic, and external pressure terms. As the parameter $\sigma_+$ decreases through the non-singular regime we have seen that we encounter first the strictly spherically symmetric and compressible flows at $\sigma_+ = 0$, and then the reversal of the flow direction around the central hyperbolic fixed point as the stress relative to its value at infinity becomes negative. In a sense then, there is a parametric topological instability in these accretion flows which permits the existence of a stable 'star' or degenerate nodal fixed central point (spherically symmetric accretion) for only one value of the transverse stress relative to infinity, namely zero.

We nominate the restricted domain of pure radial accretion in transverse stress parameter space as the origin of the 'cored apple' bipolar structure. It is this 'instability' to global convection with respect to fluctuations in the transverse stress that may account for the existence of bipolar flows around cold, low-luminosity YSOs . The existence of two modes (bipolar flow: $\sigma_+ > 0$, disc accretion: $\sigma_+ < 0$) of flow is a clear prediction of the present theory.

Admittedly, the generality of this proposition depends on our analytic model being topologically generic despite the physical restrictions stated at the beginning of §2. Consequently we have also explored a more general form of this class of stationary model wherein the velocity and density (we do not need other variables for this argument) have the forms:

$$\mathbf{v} = \sqrt{\frac{\gamma}{r}} \left(\frac{r}{r_o}\right)^{\alpha_v} \mathbf{U}(\theta), \tag{12}$$

$$\rho = \left(\frac{M_*}{r^3}\right) \left(\frac{r}{r_o}\right)^{\alpha_m} \mu(\theta), \tag{13}$$

where $r_o$ is some characteristic scale or fiducial radius. Then the continuity equation *alone* allows us to give the explicit form of the poloidal stream lines as

$$r^{\alpha_v + \alpha_m - 3/2} \mu U_\theta \sin\theta = Q \tag{14}$$

where $Q$ is a stream line constant.

Since symmetric bipolar outflow–accretion patterns require $U_\theta \to 0$ at $\theta = 0$ and $\theta = \pi/2$, we can deduce a topological instability from Eq. (14) alone. Clearly depending on the sign of $\alpha_c = \alpha_v + \alpha_m - 3/2$ we have either a central saddle point ($\alpha_c > 0$) or a central stable node ($\alpha_c < 0$). In the latter case the stream lines are closed loops which originate and end at the central node. The case $\alpha_c = 0$ requires $U_\theta = 0$ for a non-singular flow, hence just as above this case of a radially accreting (or effluxing) node is very restricted in parameter space.

The indices $\alpha_v$ and $\alpha_m$ are specified by the remaining equations of the model. In particular $\alpha_v = 0$ and $\alpha_m = 5/2$ correspond to a saddle bipolar flow with a Keplerian equatorial disc and constant dynamic viscosity. Deviations from the law $\alpha_v = 0$ require in general pressure-supported, 'virialised' thick discs as in the analytic example presented here ($\alpha_v = -1/2$, $\alpha_m = 3$). Cases where $\alpha_v > 0$ exist with self-gravity dominant and will be reported elsewhere.

Thus *the analytic solution does reflect a generic global property of this class of model*. While it can not produce realistic outflows at infinity, unlike more complicated members of the class, it does have the rare merit of being analytic.

## 3 DISCUSSION

Two typical examples of these modes are illustrated in Figures 1–3 and 4–6. We use $M_A = 1.5$ since we find that the behaviour is not sensitive to the field strength, as is to be expected in super-alfvénic flow. We have chosen $k$ such that the turbulent speed is either 10% or 3% of the free fall speed according to equation (16) below. The scale of the flow and especially the extent of the low pressure regions is rather sensitive to the value of $\ell$. We have chosen values such that the various features are clearly illustrated.

In Figure 1, we find a well-developed bipolar outflow case. The stream lines turn sharply upward in the axial low-pressure collimated region on the scale shown after they have rebounded from the equatorial pressure torus, which is mapped by the colour contours. The poloidal velocity component (Fig. 2) and the 3-D stream lines (Fig. 3) illustrate clearly that the axial stream lines have passed through the disc very close to the equatorial plane, and are a dramatic illustration of the high speed axial injection near the fiducial scale.

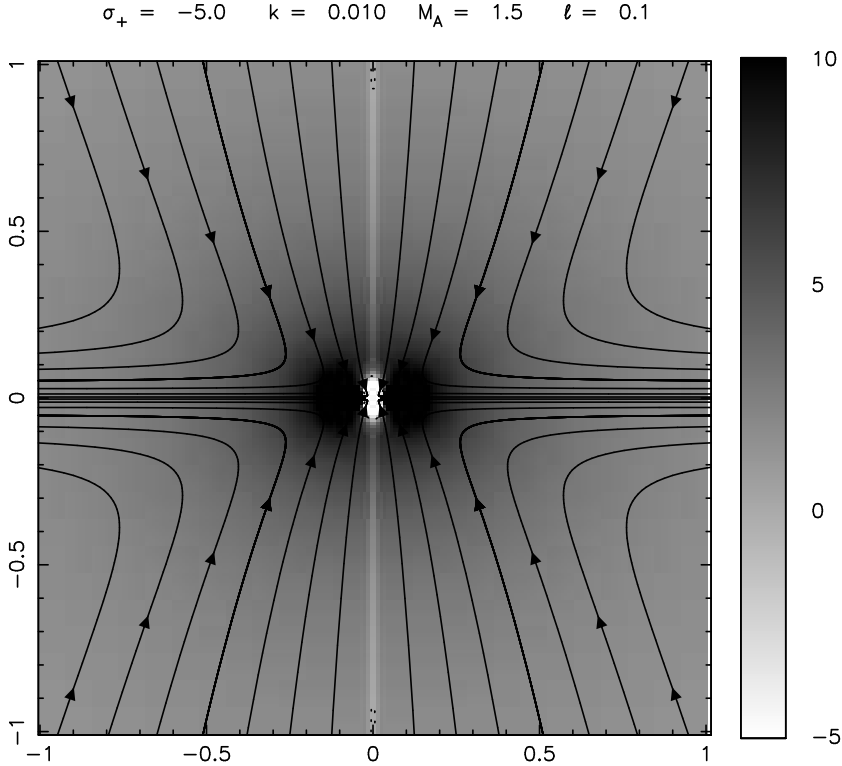

**Figure 4.** Bipolar *accretion* mode with $\sigma_+ = -5.0$, $\ell = 0.1$, $M_A = 1.5$, and $k = 0.010$. The same quantities as in Fig. 1 are represented for this case. One should note the very narrow region of negative pressure difference along the core of the 'apple', and in the high velocity region near the central source.

Figures 4–6 deal with the second mode identified in this paper. The pressure contours and the meridional stream lines on which the material now 'descends' onto the pressure disc above and below the equatorial plane are indicated in Figure 4. Now the central region is the primary low pressure high velocity region, and the stream lines are deflected either by the pressure toroid or by an equatorial high pressure zone. Figure 5 is noteworthy in that it shows two 'stagnation cones' in the velocity field which correspond to the loci of the turning points of the stream lines, while in Figure 6 we see the spiralling descent and nearly radial ejection for the stream lines which pass close to the equatorial plane.

Since the velocity falls off more rapidly than the escape velocity in both of these modes, they can only describe a small-scale circulation whose loops would be closed physically in a non self-similar region. They are thus essentially bound convective modes in the 'atmosphere'. Self-gravitating (but non-analytic) members of this class are *not* subject to this restriction.

The analytic model proposed here produces naturally and for the first time the *geometry* normally associated with bipolarity (thick discs, evacuated axial regions) in an accretion-driven circulation. It does *not* include consistently the source of free energy required to produce supra-thermal outflows at infinity. However even relatively weak radiative heating at the central source will suffice once the outflow speed there equals the free fall speed. We therefore expect our model to be most applicable to low luminosity sources, since high velocity winds and direct radiation pressure may play a rôle in more energetic sources. Future numerical studies of this type of model *can* however allow for energetic central sources.

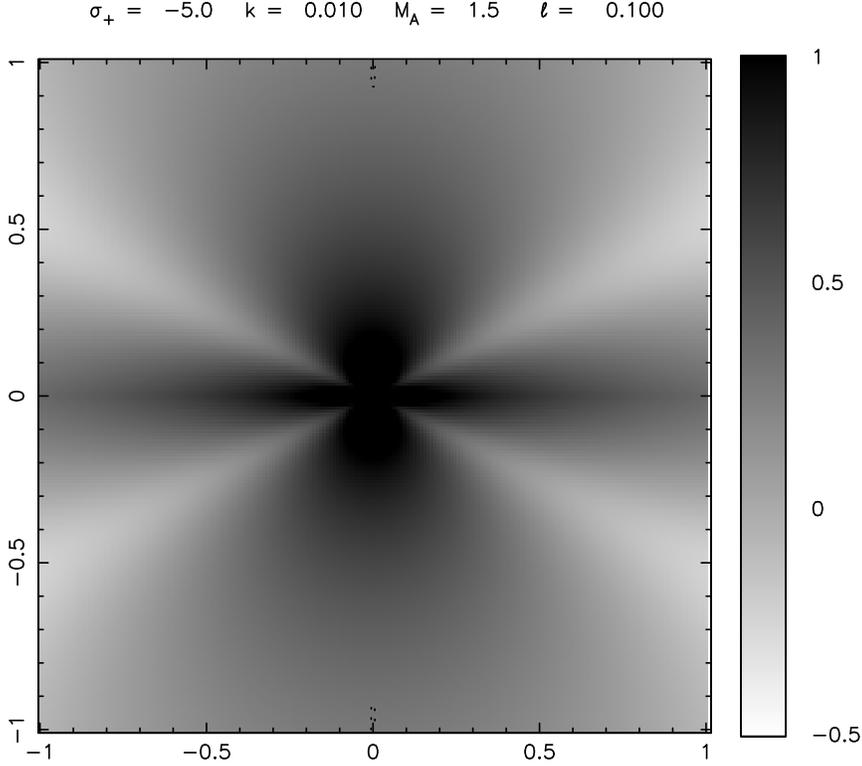

**Figure 5.** Grey-coded contours of $\log_{10}\left(v_r^2 + v_\theta^2\right)^{1/2}$ for the outflow mode of Fig. 4.

We use the uniform density isothermal interpretation of the molecular material surrounding the YSO VLA 1623 (André, Ward–Thompson and Barsony, 1993) which gives a mass of $0.6 \pm 0.3 M_\odot$, and an outer radius of $\approx 3.5 \times 10^{16}$ cm and hence a mean density of $\rho \approx 6 \times 10^{-18}$ g cm$^{-3}$. We adopt a mass $M_*$ for the central stellar core that is less than or equal to the mass of the surrounding cloud $M_c$. It is convenient to write the pressure in equation (1) as

$$\Delta \equiv \frac{p - p_\infty}{\rho \gamma / r_o} = q + akq^2 \left( P(\theta) - \frac{|\mathbf{U}|^2}{2(M_A^2 - 1)} \right) \qquad (15)$$

where $q \equiv r_o/r$, $r_o$ is a fiducial radius, and

$$ak \equiv (\nu^2/r_o^2)/(\gamma/r_o) \equiv 2v_t^2/v_{ff}^2 \qquad (16)$$

with $v_t$ and $v_{ff}$ respectively the turbulent and free fall velocities at $r_o$. Then $v_{ff}(q) = 42\sqrt{(M_*/M_\odot)/(r_o/1\,\mathrm{AU})}\,q^{1/2}$ km s$^{-1}$, and consequently, using the mean density of VLA 1623, the scale factor for the scaled pressure difference $\Delta$ is

$$\rho \gamma / r_o = 3.8 \times 10^{11} \left(\frac{M_*}{M_\odot}\right) \left(\frac{r_o}{1\,\mathrm{AU}}\right)^{-1} q \ \mathrm{K\,cm}^{-3}. \qquad (17)$$

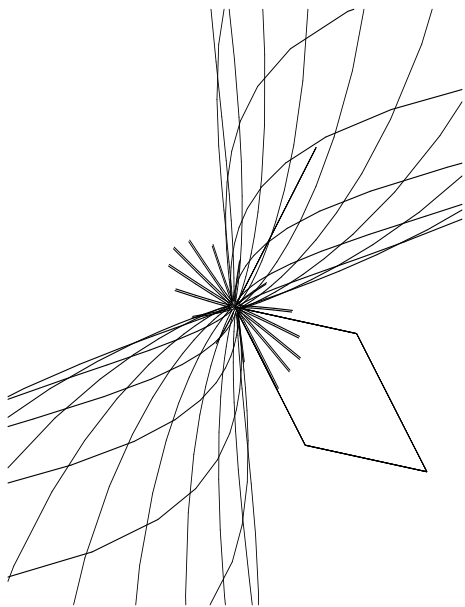

**Figure 6.** Three dimensional representations of the stream lines for the case described in Figs. 4 and 5.

This scale is to be used with the pressure contours in all the figures. With stellar core temperatures of a few thousand degrees, $\Delta = 1$ implies a density of $O(10^8)$ cm$^{-3}$.

From equation (1) we may now write for the parameters displayed in Figure 1 the velocity scale as

$$\mathbf{v} = 0.4 \left(\frac{M_*}{M_\odot}\right)^{1/2} \left(\frac{r_o}{1\,\mathrm{AU}}\right)^{-1/2} q\,\mathbf{U}\ \mathrm{km\,s}^{-1}, \tag{18}$$

and the magnetic field scale as

$$\mathbf{B} = 7 \times 10^{-4} \left(\frac{M_*}{M_\odot}\right)^{1/2} \left(\frac{r_o}{1\,\mathrm{AU}}\right)^{-1/2} q\,\mathbf{U}\ \mathrm{Gauss}\,. \tag{19}$$

The limitations of this analytic model for the discussion of real bipolar outflows are now apparent. For instance (Figure 1) we find that in the equatorial plane $\mathbf{U} = (-7.5, 0, 10)$, while at an angle of $8°$ to the polar axis $\mathbf{U} \approx (100, -12, 71)$ at $q = 1$. This latter point is well inside the region where the pressure is less than that at infinity for $\ell = 10$, but by reducing this to say 2 it would not be so. But even though velocities characteristic of the flow in VLA 1623 ($\approx 50 \pm 25$ km s$^{-1}$) can be produced and well collimated at 1 AU, the velocities will be reduced to nothing at 100 to 1000 AU. Yet it should be observed that much has been gained already simply by redirecting the free fall of the gas outward along the axis so that most of it misses the central source. Not only is the dissipation of all this energy avoided so that a low luminosity is attained, but, in addition, by moderately heating this gas near the centre, the flow will persist at infinity. By Bernoulli's theorem, the terminal speed $v_\infty = \sqrt{2h}$, where $h = (\gamma/\gamma - 1)(kT_*/\bar{m})$ is the specific enthalpy, can reach $v_\infty \sim 21$ km s$^{-1}$ for a gas temperature near the star of $T_* = 10^4$ K, which is compatible with the observations of the central source. An intrinsic way of producing high velocities also seems possible in self-gravitating flow with $\alpha_v > 0$ as remarked above. This seems to be due to distributing the retarding gravitational mass throughout the protostellar flow.

It is interesting to note that higher velocity material will be injected into the funnels at smaller radii and so will overtake the outer material. Should, as a result of expansion, the pressure of the material in the funnel drop to zero, then the material injected at different heights in the funnel can interpenetrate. Then we have the basis for a 'Milne' model of bipolar lobes, wherein material launched with a range of subsequently constant velocities rearranges itself into a Hubble flow (eg. Shu et al., 1991; Masson and Chernin, 1992).

In summary, we have proposed a model of bipolarity which is a global circulation or convection exterior to a gravitating core. There are two equally available modes in the steady solutions which describe respectively an axial bipolar outflow driven by an equatorial pressure torus, and an axial accretion onto this standing toroidal pressure wave. This latter mode is a distinct testable prediction of the theory, since this behaviour is the natural development from a generic topological instability in a purely radial accretion flow.

**ACKNOWLEDGEMENTS**

We warmly thank Philippe André for constructive remarks. This work was supported by a grant from the Canadian NSERC.